\documentclass{sig-alternate}
\usepackage[utf8]{inputenc}
\usepackage[T1]{fontenc}
\usepackage{microtype}

\usepackage{hyperref}
\usepackage{listings}
\usepackage{graphicx}
\usepackage{paralist}
\usepackage{tabularx}

\usepackage{flushend}

\begin{document}
\conferenceinfo{RCoSE}{'14, June 3, 2014, Hyderabad, India}
\CopyrightYear{14}
\crdata{978-1-4503-2856-2/14/06}

\title{Scrum for Cyber-Physical Systems:\\ A Process Proposal}

\numberofauthors{1} 
%
%
%
%
%
\author{
\alignauthor
    Stefan Wagner\\
    \affaddr{University of Stuttgart}\\
    \affaddr{Institute for Software Technology}\\
    \affaddr{Stuttgart, Germany}\\
    \email{stefan.wagner@informatik.uni-stuttgart.de}}

\maketitle
\begin{abstract}
Agile development processes and especially Scrum are changing the state of the practice
in software development. Many companies in the classical IT sector have adopted them
to successfully tackle various challenges from the rapidly changing environments and increasingly
complex software systems. Companies developing software for embedded or cyber-physical
systems, however, are still hesitant to adopt such processes.

Despite successful applications of Scrum and other agile methods for cyber-physical systems,
there is still no complete process that maps their specific challenges to practices
in Scrum. We propose to fill this gap by treating all design artefacts in such a development in the
same way: In software development, the final design is already the product, in hardware and
mechanics it is the starting point of production.

We sketch the Scrum extension Scrum CPS by showing how Scrum could be used to develop all 
design artefacts for a cyber physical system. Hardware and mechanical parts that might not be
available yet are simulated. With this approach, we can directly and iteratively build the final software
and produce detailed models for the hardware and mechanics production in parallel.

We plan to further detail Scrum CPS and apply it first in a series of student projects to gather
more experience before testing it in an industrial case study.

\end{abstract}




\section{Introduction}

Embedded software systems, mobile and desktop applications as well as Internet and cloud
systems are converging into large, complex and distributed systems that interact with the
real world via sensors and actuators. For this new kind of systems, the term \emph{cyber-physical
systems} has been coined. ``Cyber-physical systems (CPS) are physical and engineered systems 
whose operations are monitored, coordinated, controlled and integrated by a computing and 
communication core.''~\cite{Rajkumar:2010hk}

Building these complex systems is still a challenge. For example, good software engineering processes 
are essential to get reliable CPS~\cite{Lee:2006vf}. In the classical information system development,
agile approaches have gained a large acceptance and provide many advantages beyond traditional
approaches. Even for embedded software, agile development is more and more accepted.
Salo and Abrahmsson~\cite{Salo:2008hc} conclude from a survey with European embedded software
development organisations that ``The results also indicate that the appreciation of the agile methods 
and their individual practices appears to increase once adopted and applied in practice.''

\subsection{Problem Statement}
Yet, outside of software development, agile development process are not widely used. ``Agile system 
engineering practices have matured for software projects while hardware system engineering continues 
to embrace classical development techniques.''~\cite{Huang:2012er} A general barrier for using agile methods
for hardware development is the higher difficulty in modifying hardware. In addition,  there is
no clearly defined, detailed description how Scrum should be used for hardware development and
how the integration with software development should work which is essential for CPS.
Lee~\cite{Lee:2006vf} describes challenges and research directions in cyber-physical systems and
proposes, among others, to ``Rethink the hardware/software split''. 

\subsection{Research Objectives}
We have the general goal to provide a clearly described, detailed and applicable Scrum variation for 
CPS development called \emph{Scrum CPS} together with empirical evidence that it works for 
developing such systems. Scrum CPS shall give concrete guidelines how to build all parts of a CPS
and ensure their integration. In this paper, we aim to provide a first sketch of Scrum CPS for further
discussions.

\vspace{3em}
\subsection{Contribution}
This paper is only a first step to reach our objectives. We proposes a first draft of Scrum CPS
based on a discussion of CPS challenges and existing work. Its innovation is the explicit handling of 
concurrent hardware and software development
and a concentration on building explicit design models for hardware components that can be simulated
together with the development to enable early software- and hardware-in-the-loop tests.

\section{Related Work}

The original way to manage more than one Scrum team working on the same product is the
\emph{Scrum of Scrums}~\cite{rubin13}. It is an additional Scrum on top of the other teams composed of members
of each of these teams. They are chosen in a way so that they can best discuss the inter-team
dependencies. This approach, however, does not prescribe any common and fixed synchronisation
points which are necessary for the complex interplay of different hardware/software components
in an CPS.

Leffingwell~\cite{leffingwell11} introduces the concept of an \emph{Agile Release Train} (ART) as a
metaphor for synchronising agile teams. In certain intervals (\emph{cadence}) the Scrum teams working
in parallel have to put their increments on the release train to create a (potential) release. These
intervals comprise several sprints. A team can also decide to put no new features on the train and
only adapt their interfaces to the changes made by other teams and their features and components.
We see this as a good fit to the diverse components needed to be developed for a CPS and employ the
ART in our proposed process.

Xie et al.~\cite{Xie:2012fz} describe a preliminary systematic review on empirical studies of the use
of agile methods in embedded software development. They describe relevant characteristics of
embedded software development such as hardware dependence or specific development environments.
We take these characteristics and discussed responses into account but the review is rather brief.

Shen et al.~\cite{Shen:2012ku} provide a more comprehensive systematic review on agile in embedded
software development. They give a good overview on the literature describing the experiences so far.
They conclude that the state of theory as well as ``research on applying agile methods to embedded
software development is distinctly not mature |\ldots]''. We describe a theoretical proposal in this paper
which we plan to flesh out in more detail and then empirically evaluate in future projects. 

Srinivasan, Dobrin and Lundqvist~\cite{Srinivasan:2009vq} also describe the state of the art in agile
for embedded systems development. They concentrate more on the organisational change aspect of
introducing such methods while we propose a concrete process.

Similarly, Cawley, Wang and Richardson~\cite{Cawley:2010fi} discuss the state of the art in using
lean and agile methods in regulated, safety-critical systems. They conclude that corresponding standards
can be mapped to agile practices although not always satisfactorily. Further issues are discussed, for 
example that refactoring can invalidate earlier certification. They state that ``It would be useful to look 
at the governance of Lean/Agile software development in these domains with a view to identifying 
how to design policies and product lifecycles [\ldots]''. W react to this challenge with an explicit process
variation for safety-critical systems.

Huang, Darrin and Knuth~\cite{Huang:2012er} describe positive experiences with using Scrum and
XP practices in systems engineering for satellite development. They stress that development phases
had to change but describe not in detail how different Scrums are used to achieve this. They give,
for example, the closeness to the sponsors as a positive effect of agile system engineering.

Cordeiro et al.~\cite{Cordeiro:2008fx} give the most detailed agile methodology for developing
embedded software. They use some practices of XP and Scrum to achieve a highly iterative process.
We will build on their work, concentrate on clear guidelines for applying Scrum and extend it to
CPS.

\newcolumntype{R}[1]{>{\raggedleft\arraybackslash}p{#1}}
\renewcommand{\arraystretch}{2}
\begin{table}[!htbp]
  \caption{Major Process Challenges and Solutions in Scrum CPS (adapted and extended from~\cite{Rajkumar:2010hk})}
  \centering
  \begin{tabular}{R{1.9cm}p{5.5cm}}
    \hline
    \textbf{Software/ hardware co-design} & CPS consist potentially of new software, electrical, electronic and 
    		mechanical components. They all need their own timelines for development but need to be synchronised.
		We use the Agile Release Train as method to structure the different Scrum teams. In addition, we define
		always a major design team which initially and regularly looks at the overall CPS architecture.\\
    \hline
    \textbf{Robustness, safety, security} & Early and comprehensive design using software directly as well as
    		design models that we can simulate help to detect quality problems early in software- and hardware-in-the-loop tests.\\
    \hline
    \textbf{Architecture} & By using detailed design models together with simulations we build early consistent and
    		global architectures. Bringing the design models into the sprint backlog of hardware sprints, we ensure
		that all components reflect this architecture.\\
    \hline
    \textbf{Real-time abstractions} & Again, the early design models can reveal bottlenecks and problems with
    		real-time constraints. In addition, the possibility of early prototypes and hardware-in-the-loop tests
		can be used to assure such constraints.\\
    \hline
    \textbf{Sensor and mobile networks} & CPS emphasise the connectedness of many quite different components. To
    		some degree, the availability of models for simulations and in-the-loop tests helps to test these aspects. In
		addition, the Agile Release Train ensures compatible interfaces between components connected via networks.\\
    \hline
    \textbf{Model-based development} & For everything but software directly, Scrum CPS demands model-based development.
    		We always build models we can simulate together for all components including the created software.\\
    \hline
    \textbf{Verification, validation, certification} & Scrum CPS enforces an early verification and allows the developers to
    		regularly and comparably early validate the system with stakeholders. Certification is more problematic and needs
		a mapping of the applicable standard to Scrum CPS.\\
    \hline
  \end{tabular}
  \label{tab:challenges}
\end{table}

\begin{figure*}[!htbp]
\begin{center}
\includegraphics[width=.7\textwidth]{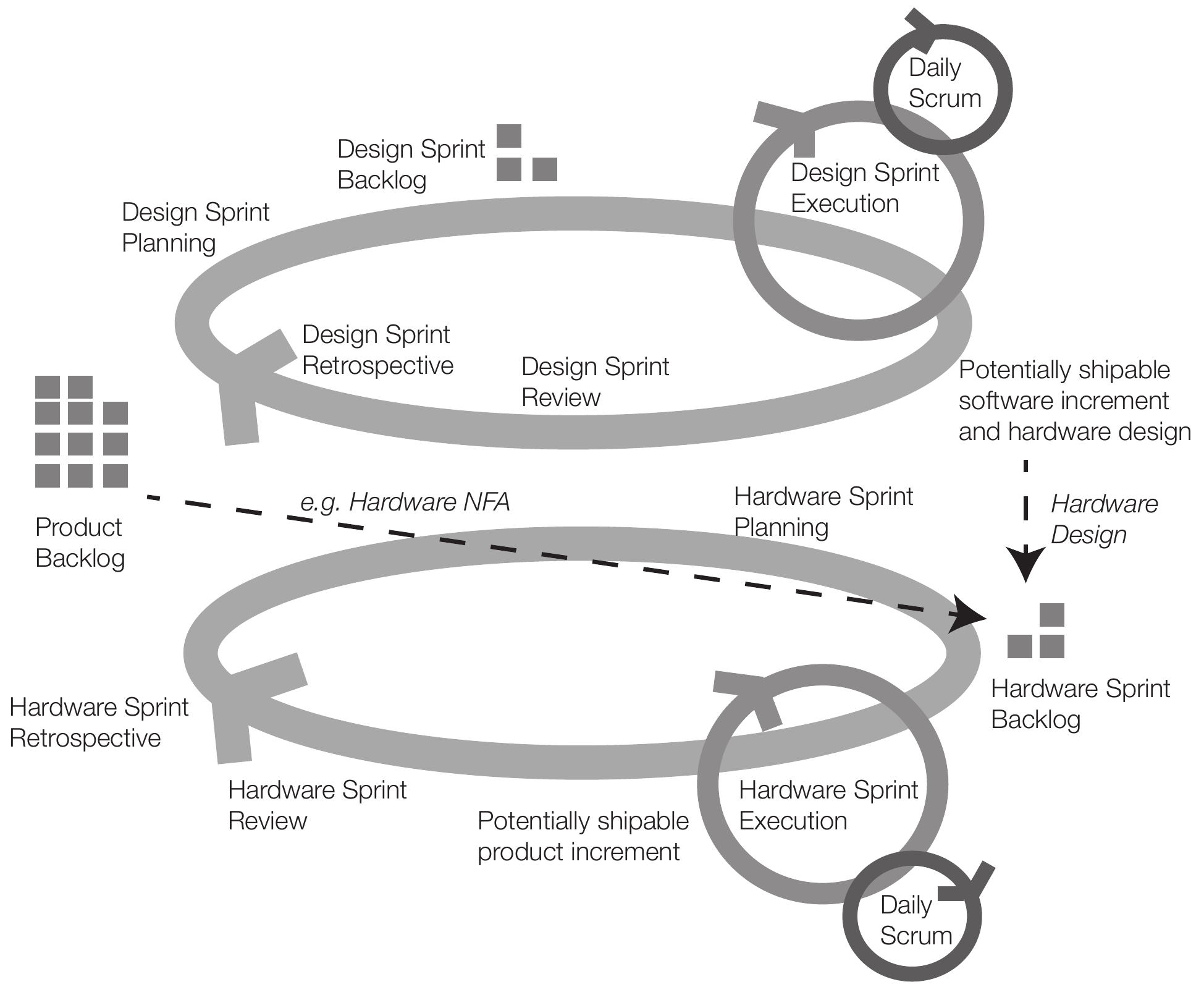}
\caption{An overview of the Scrum CPS process}
\label{fig:process-sketch}
\end{center}
\end{figure*}

\section{Scrum CPS}

We propose the Scrum variation called \emph{Scrum CPS} as a clearly described and defined process
for the agile development of cyber-physical systems. To address the real needs in developing CPS, we
first look at the challenges posed by CPS development onto the development process. We summarise
the major challenges with our proposed solutions in Tab.~\ref{tab:challenges}.

The central idea of the Scrum CPS process is sketched in Fig.~\ref{fig:process-sketch}. We designate a sprint either as
\emph{design sprint} or \emph{hardware sprint}. The initial sprints are always design sprints starting with central parts of
the architecture of the CPS. As software is design and product at the same time, we can already build operational software.
This software is developed and tested together with hardware simulations from design models we build in
parallel.

The produced hardware designs (together with potentially other items from the product backlog) become part of the sprint
backlog for the subsequent hardware sprint. The hardware sprint refines and extends the design models and builds the
actual hardware. First we produce prototypes and later the blueprints for the final hardware. At the end of the hardware sprints, we get
potentially shippable product increments as well as modifications for the product backlog which can then be introduced
into design sprints. A product increment is then really a combination of software and hardware that could be given to a
user -- although the hardware might still be crude, it would be useable. 

\begin{figure}[htbp]
\begin{center}
\includegraphics[width=.5\textwidth]{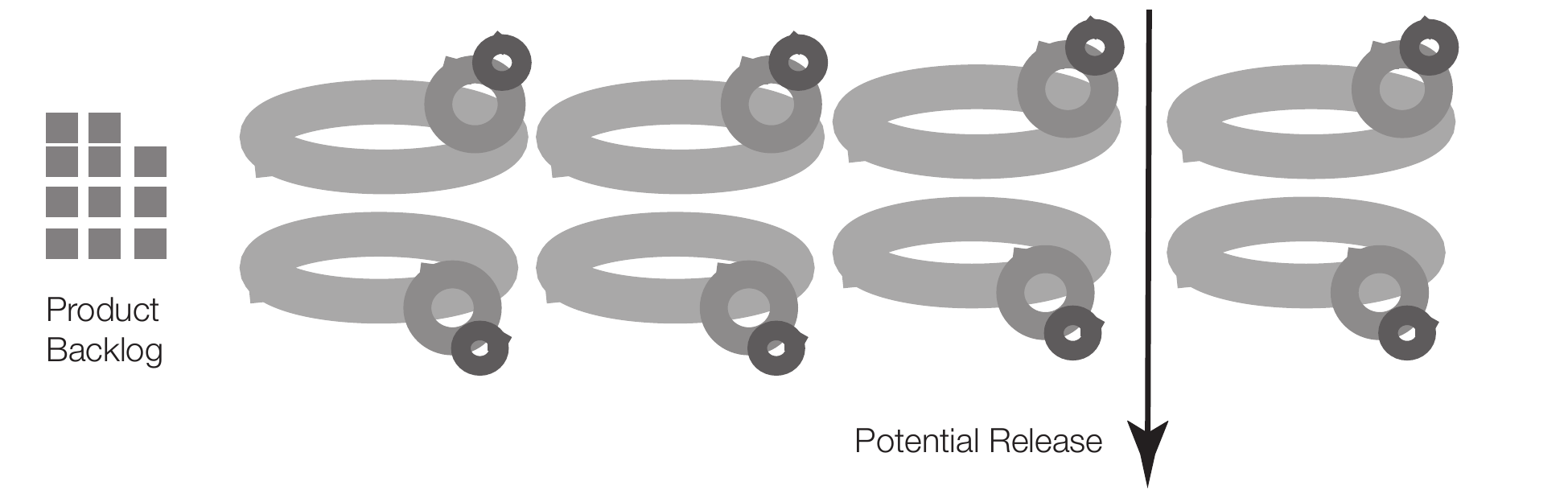}
\caption{The Agile Release Train}
\label{fig:release-train}
\end{center}
\end{figure}

To ensure that these different sprints are synchronised and produce a coherent CPS, we adopt the ida of an Agile Release
Train (ART) from \cite{leffingwell11} as shown in Fig.~\ref{fig:release-train}. The whole execution of the process is structured
into cascades with a potential release at the end. Before each release, there can be several sprints by each team. Yet, any
team knows that they have to deliver something for the release. In the worst case, they will deliver the last version with possibly
adapted interfaces only.

In the following, we discuss the two types of sprints in more detail and propose variations depending on the concrete
application area of the CPS.

\subsection{Design Sprints}

Design sprints are responsible for the system design. They produce designs for hardware and software components. While for
hardware this means an executable description of an abstraction of the future hardware, the design for software is the software
itself. We often differentiate in software engineering between design and implementation. Design then contains architecture
descriptions with the proposed decomposition into components or interface descriptions. As in Scrum in general, this will be
used and created in Scrum CPS as well where appropriate. Yet, in comparison to hardware, software itself is still a design
because there is no manufacturing step in the process. Hence, we can see software itself on the same level as detailed,
behavioural hardware designs. 

This means, we produce workable software together with hardware designs we can simulate and, thereby, ``integrate'' the
software and hardware designs in early design sprints. In some sense, because working software is created first, we could
call this \emph{software-driven system engineering}. Using software and hardware designs, we can -- fitting to usual agile
ways of development -- test from the first sprint on using software-in-the-loop tests.

The hardware could be designed, for example, with the rapid hardware definition language (Rapid HDL) as proposed 
in \cite{Allen:2009hz}. It allows to script hardware using ``reusable software objects, communication between hardware 
and software is automatic, and synthesis is automated using a free tool chain.'' This then produces a suitable basis for the
sprint backlog of further design sprints as well as hardware sprints.

Further challenges, we have not yet explicitly incorporated into the design sprint process are how to handle the complexities
of highly networked and real-time components. We will need to work on that in more detail to understand whether it can
be addressed by the process itself or if it is a matter of the modelling and development methods used.

\subsection{Hardware Sprints}

The result of the design sprints are ``only'' design models we can simulate. This is already a very good basis for further
hardware development as the models show how hardware and software will need to interact. Yet, there are many more
detailed decisions to be made in hardware development. This depends to a large degree on what kind of hardware we
need to build. It also might make sense to structure the hardware development into separate Scrum teams working on
electrics, electronics and mechanics, for example. 

Based on the design models and further detailing, the hardware engineers build prototypes which show certain aspects
of the future hardware. This includes laying out the hardware designs and creating a bill of materials. Then we order the
prototypes or assemble them ourselves. Programmable logical devices can play the role of quickly adaptable hardware.
This was, for example, already recognised in the development of the first Macintosh at Apple: ``Burrell Smith developed a 
unique hardware design style based on programmable array logic chips (PAL chips), which enabled him to make changes 
much faster than traditional techniques allowed, almost with the fluidity of software.''\footnote{\url{http://www.folklore.org/StoryView.py?project=Macintosh&story=The_Macintosh_Spirit.txt}} 
   Hence, we can now test important aspects to refine the hardware designs and 
add additional or changed
items to the product backlog so they can be worked on in the next design sprints. Depending on the nature of the
prototypes, we aim to run hardware-in-the-loop tests with the prototypes and the existing software from the design sprints.

As soon as any hardware prototypes are available, the sprint will contain inspections, such as visual inspections,
multi-meter checks and turn-on tests. Based on these inspections and tests, we probably need to debug the prototype.
This will lead to changes in the design for the design sprints.

After several hardware sprints and a synchronisation with the design sprints that the design is stable, we will take
the final steps to mass produce the hardware.

\subsection{Variations}

In the following, we will discuss some potential variations of Scrum CPS depending on the nature of the components
of the system.

\subsubsection{Safety-Critical Components}

In many cases, because of the close connection to the physical world, at least parts of an CPS are  safety-critical.
Safety-critical means here that the component can create hazards for the system environment that can lead to accidents
harming people or with other catastrophic consequences. These components and the system containing these components need 
to be developed with special care to avoid these consequences. In particular, the development organisation needs to follow
applicable standards for the corresponding domain of the component. For electronic and software components, the IEC 61508
needs to be followed. Certain domains have created specific standards, such as the ISO 26262 for such components in
automotive systems. Those standards usually prescribe various aspects of the process, sometimes also specific methods
and product properties. Depending on the standard, we might need to modify Scrum CPS to conform with it. Some aspects 
are easy to achieve or at least not more difficult in a Scrum setting. For example, MC/DC test coverage or traceability links from
requirements to code is often required for higher safety integrity levels. These would be good candidates for inclusion into
the \emph{definition of done} of a sprint.

\subsubsection{New Cloud Components}

In the design sprints, we assumed so far that we only build design models for the hardware components but build the concrete software
components as in the final product. A variation could be used for the more cloud-dependent parts of the CPS. Maybe the cloud part
is not yet available or we do not want to reveal it yet. Then it might be acceptable to only build a simulation of the cloud
service.

\subsubsection{Hardware Design Language}

Scrum CPS assumes that the hardware design is always an executable simulation. Scrum CPS does not prescribe what modelling language
should be used for the simulation. This could simply be the programming language used to build the other, productive software
over general purpose modelling languages, such as SysML with corresponding tools, to specialised modelling languages
and tools, such as Matlab Simulink and Stateflow. This might change the process and when and how it is able to run software-
and hardware-in-the-loop tests.

\section{Example: Fitness Tracking}

\begin{figure}[htbp]
\begin{center}
\includegraphics[width=.4\textwidth]{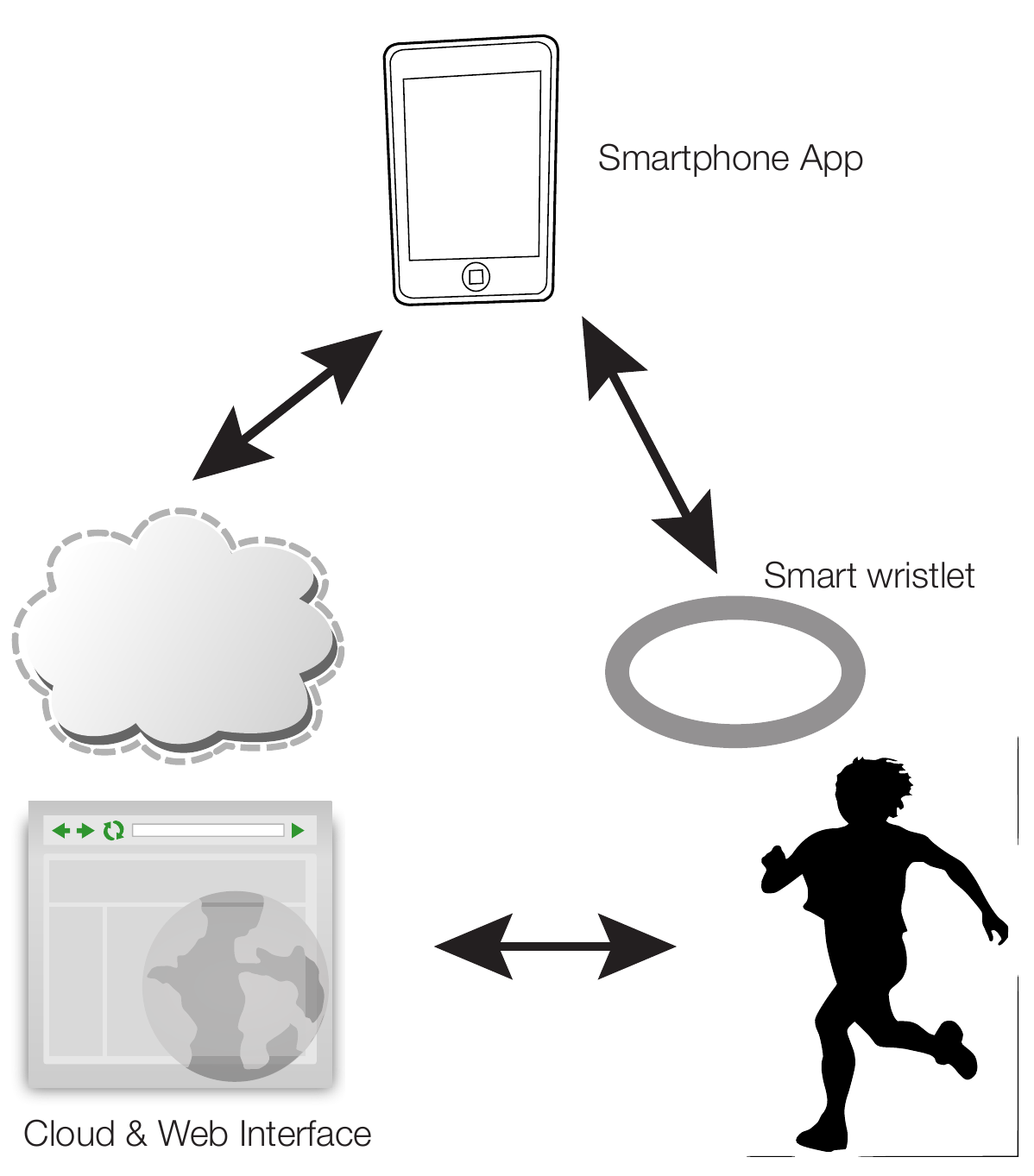}
\caption{Overview of the Fitness Tracking CPS}
\label{fig:example}
\end{center}
\end{figure}

To make the process proposal more concrete, we apply it to a hypothetical fitness tracking CPS. The purpose of the system
is to track the movements of a person and to give feedback to motivate her or him to exercise more. The system consists
of a \emph{smart wristlet} with sensors, actuators and a small screen, a smartphone app on a modern smartphone and a
cloud service with a corresponding web interface. This structure is shown in Fig.~\ref{fig:example}.

The wristlet tracks the movements of the person, connects via Bluetooth to the smartphone for syncing the movement
data and shows a movement score. With the synced information, it also vibrates when more movement is needed, and it
suggests activity via the small screen. The smartphone app connects to the cloud service via Internet. The cloud service
captures all the movement data and personal configurations. It also allows social functions such as comparing one's results
with those of friends.

We chose this example because it is a well-known type of system and it contains only a small portion of hardware development.
This limits the representativeness of the examples but allows a more thorough discussion. We assume we only have to
develop the wristlet hardware. All smartphone and server hardware is standard.

Following Scrum CPS, we start with creating a product backlog. We cannot go into details of the backlog but discuss
some examples. In general, the above mentioned vision should be adequately represented. Example backlog items
will be more general, design-oriented, such as ``As a user, I want to be able to see my movement data in numerical
and diagrammatical form on my smartphone.'', as well as more concrete software or hardware related items, such as
``As a user, I want to be able to connect my wristlet to my smartphone.''

Let us assume we have two teams of about 10 people available to build our CPS. As we have hardware development 
involved, we go for a 90-days cadence with 30-days sprints. Next, we plan the first sprints which will be design sprints 
for both teams. As the biggest innovation and value for the product owner will be in the wristlet itself, one team starts 
with a design sprint for it. This fits also to the expectation that there will be more hardware sprints needed to develop
the wristlet. An early concentration on its design helps to avoid waiting for the hardware later. The other team chooses the
second highest area of value: the smartphone app. Both teams select the corresponding backlog item for their sprint 
backlogs and start the sprint executions. In the executions, we follow the regular Scrum practices like Daily Scrums. 
We build the hardware design as CAD models and in Rapid HDL. In parallel, we start with implementing the first stories 
in software to be run on the hardware to be developed and the smartphone. We close the sprint with software-in-the-loop tests.

In the next sprint planning, we decide to continue with design sprints for both teams as we have several more design-related
backlog items for the wristlet. We select them into the sprint backlog for one team and smartphone app items in the backlog for
the other team. We execute the sprint as above but decide after the tests that the hardware design is ready to be built as a 
prototype. Hence, in the following sprints, we have one hardware sprint for the wristlet and a design (software) sprint for
the smartphone app. In the hardware sprint, we build up the first hardware prototype and improve it while running the existing software
on it. In the design sprint, we extend and improve the corresponding software. We can now run also 
hardware-in-the-loop tests to assure the quality of the product increment. Both sprints also are allowed to give feedback
for changing the product backlog based on the experiences of the sprint.

After these three sprints, we have reached our first synchronisation point where each team has to put its cargo onto
the release train. This means, we have a first version of the CPS we can show to (potential) customers or marketing and
get feedback. As we still have prototype hardware, we will not make a full release out of it, however.

In the next sprints, we extend the system sprint by sprint by the smartphone app and the cloud service. In the hardware sprints, we iteratively
improve the hardware and finally give it to a full assembly. The last sprints implement the last missing backlog
items and concentrate on system testing the whole CPS.

\section{Conclusions}

In this paper, we have proposed a variation of Scrum called \emph{Scrum CPS} for a clear definition of how to use Scrum
for developing cyber-physical systems to gain all the benefits that agile development has to offer for software-only developments.
There is a series of challenges in CPS in general and several of them also influence the needed contents of a corresponding
development process. We have briefly discussed these challenges and mapped them to our process proposal. Using a
hypothetical example of a fitness tracker CPS, we discussed the application of the process.

This brief sketch can only be a start. We are in the process of further detailing Scrum CPS to make it usable in
real CPS projects. Using this description, we plan to work with other research groups in electrical and mechanical engineering
to apply Scrum CPS in student projects. We will closely monitor these projects to better understand its benefits and problems
to further improve it. With the improved process description, we aim to run case studies in industry for further validate the approach.

\section{Acknowledgements}
I would like to thank Jan-Peter Ostberg who gave valuable feedback on a first draft of this paper.

\bibliographystyle{abbrv}
\bibliography{EmbeddedScrum} 

\end{document}